# New financial ratios based on the compositional data methodology

Salvador Linares-Mustarós [a, *], Maria Àngels Farreras-Noguer [a],
Núria Arimany-Serrat [b], Germà Coenders [c]

[a] *Department of Business Administration, University of Girona, Carrer Universitat de Girona, 10, 17071, Girona, Spain*
[b] *Department of Economics and Business, University of Vic-Central University of Catalonia, Carrer de la Sagrada Família, 7, 08500 Vic, Barcelona*
[c] *Department of Economics, University of Girona, Carrer Universitat de Girona, 10, 17071, Girona, Spain*

**Abstract:** Due to their type of mathematical construction, the use of standard financial ratios in studies analysing the financial health of a group of firms leads to a series of statistical problems that can invalidate the results obtained. These problems are originated by the asymmetry of financial ratios. The present article justifies the use of a new methodology using compositional data (CoDa) to analyse the financial statements of a sector, improving analyses using conventional ratios since the new methodology enables statistical techniques to be applied without encountering any serious drawbacks such as skewness and outliers, and without the results depending on the arbitrary choice as to which of the accounting figures is the numerator of the ratio and which is the denominator. An example with data of the wine sector is provided. The results show that when using CoDa, outliers and skewness are much reduced and results are invariant to numerator and denominator permutation.

## 1. Introduction

While standard financial ratios help to accurately evaluate the financial statements of firms and corporations at an individual level (Whittington, 1980; Barnes, 1987, Bernstein, 1993; Gallizo, 2005), unfortunately when they are used as variables in statistical analyses of the financial health of a sector, the reliability of the results of these diagnoses cannot be accepted as valid given that, as this study will show, the loss of symmetry produced when defining standard financial ratios causes potentially significant distortions when analysing financial health. Although this asymmetry in financial ratios has been known for some time (Lev and Sunder, 1979; Cowen and Hoffer 1982; Mcleay and Omar, 2000), it has not received due attention from the area of accounting, despite the serious problem of it casting doubt on the results of multiple current studies that use standard ratios as variables (Fernández-Olmosá et al., 2009; Hammervoll et al., 2014; Newton et al., 2015; Delord et al., 2015; Lorenzo et al., 2018). The statistical problems of ratios have also been reported in other

* Corresponding author.
E-mail addresses: salvador.linares@udg.edu (S. Linares-Mustarós), angels.farreras@udg.edu (M.A. Farreras-Noguer), nuria.arimany@uvic.cat (N. Arimany-Serrat), germa.coenders@udg.edu (G. Coenders)

scientific fields (Isles, 2020). The present article aims to correct this fact by providing a detailed explanation of the reason for this problem, which is also the root cause of other identified and related problems such as the emergence of spurious outliers (So, 1987; Watson, 1990; Creixans–Tenas et al., 2019), and by showing how the results obtained are incoherent if ratios where the numerator and the denominator are permuted are used (Frecka and Hopwood, 1983; Linares–Mustarós et al., 2018; Creixans-Tenas et al., 2019).

In the first part of this article, a mathematic counterexample will be used to evidence the serious drawbacks of using standard financial ratios in statistical studies of economic sectors. Once the danger of incurring serious methodological problems when using standard financial ratios in this type of study has been explained, a new type of financial ratios based on the methodology of compositional data analysis, or simply compositional data (CoDa), will be suggested, the validity of the results of which has already been extensively tested in other fields (Aitchison, 1986; Pawlowsky-Glahn and Buccianti, 2011; Van den Boogaart and Tolosana-Delgado, 2013; Pawlowsky-Glahn et al., 2015; Filzmoser et al., 2018; Greenacre, 2018). While the CoDa methodology emerged from the fields of geometry and chemistry at the end of the last century (Aitchison, 1982; 1986), it has since been extended to all the other scientific fields of study, including economics and other social sciences (Coenders and Ferrer-Rosell, 2020), and has started to be regularly used in studies in the area of finance (Voltes-Dorta et al., 2014; Ortells et al., 2016; Belles-Sampera et al., 2016; Davis et al., 2017; Verbelen et al., 2018; Boonen et al., 2019; Kokoszka et al., 2019; Wang et al., 2019; Gámez-Velázquez and Coenders, 2020; Maldonado et al., 2021a; 2021b; Porro, 2022; Vega-Baquero and Santolino, 2022), and, more recently, in the area of accounting (Linares-Mustarós et al., 2018; Creixans-Tenas et al., 2019; Carreras-Simó and Coenders, 2020; 2021; Saus-Sala et al., 2021; Arimany-Serrat et al., 2022). The present article aims to show that the CoDa approach can be used in the accounting field ensuring the validity of results, and to provide the research community in finance and accounting with a reasoned case for using a new methodology to analyse the financial statements in a sector. To achieve this end, the study concludes with a simple example of an analysis of the financial statements of the wine sector in a European country, which shows the invalidity of the results obtained using the traditional methodology and how the proposed methodology avoids the abovementioned problems.

Based on the line of argument presented, this study is organised in three main sections. Section 2 focuses on showing the serious problems arising from the use of standard ratios in sectoral studies, which result from the lack of symmetry of ratios and can lead to the invalidity of the results of the analysis. Financial ratios based on the CoDa methodology are presented in section 3 as possible candidates to replace standard or conventional financial ratios. Section 4 presents a research example in the wine sector comparing the use of conventional and compositional ratios, which aims to show the need to change from the usual working methodology in sectoral studies based on standard financial ratios by exposing the significant discrepancies in the two sets of results. Sections 5 and 6 present the discussion and conclusions, respectively.

## 2. Theory. The problem of the asymmetry of standard ratios. The appearance of spurious outliers

This section explains in detail why methodological works that use standard financial ratios as variables in sectoral statistical analyses cannot be considered as valid. To this end, the starting point is the fictitious data of a group of ten firms, the values of two accounting magnitudes of which are given in the first two columns of Table 1.

Table 1.
Artificial data of a ten-firm sector.

| | Magnitude 1 | Magnitude 2 | α | $\frac{\text{Mg 2}}{\text{Mg 1}}$ | $\frac{\text{Mg 1}}{\text{Mg 2}}$ |
|---|---|---|---|---|---|
| Firm 1 | 0.5 | 4 | ~82.875 (=45+37.875) | 8 | 0.125 |
| Firm 2 | 1.5 | 3 | ~63.435 (=45+18.435) | 2 | 0.5 |
| Firm 3 | 1.5 | 2.5 | ~59.035 (=45+14.035) | $1.\hat{6}$ | 0.6 |
| Firm 4 | 1.8 | 3 | ~59.035 (=45+14.035) | $1.\hat{6}$ | 0.6 |
| Firm 5 | 1.5 | 1.5 | ~45 (=45+0) | 1 | 1 |
| Firm 6 | 3 | 3 | ~45 (=45-0) | 1 | 1 |
| Firm 7 | 3 | 1.8 | ~30.965 (=45-14.035) | 0.6 | $1.\hat{6}$ |
| Firm 8 | 2.5 | 1.5 | ~30.965 (=45-14.035) | 0.6 | $1.\hat{6}$ |
| Firm 9 | 3 | 1.5 | ~26.565 (=45-18.435) | 0.5 | 2 |
| Firm 10 | 4 | 0.5 | ~7.125 (=45-37.875) | 0.125 | 8 |

Given that this section focuses on the problems related to asymmetry, the data Magnitude 1 and Magnitude 2 in Table 1 are represented graphically in Cartesian coordinates. A look at Figure 1 shows that the magnitudes recorded in Table 1 are chosen because they are symmetrical with respect to the 45° angle in the computation of the angles α of the rays that start at the origin and pass through the points, as shown in Figure 2. This symmetry is reflected in the column headed α in Table 1.

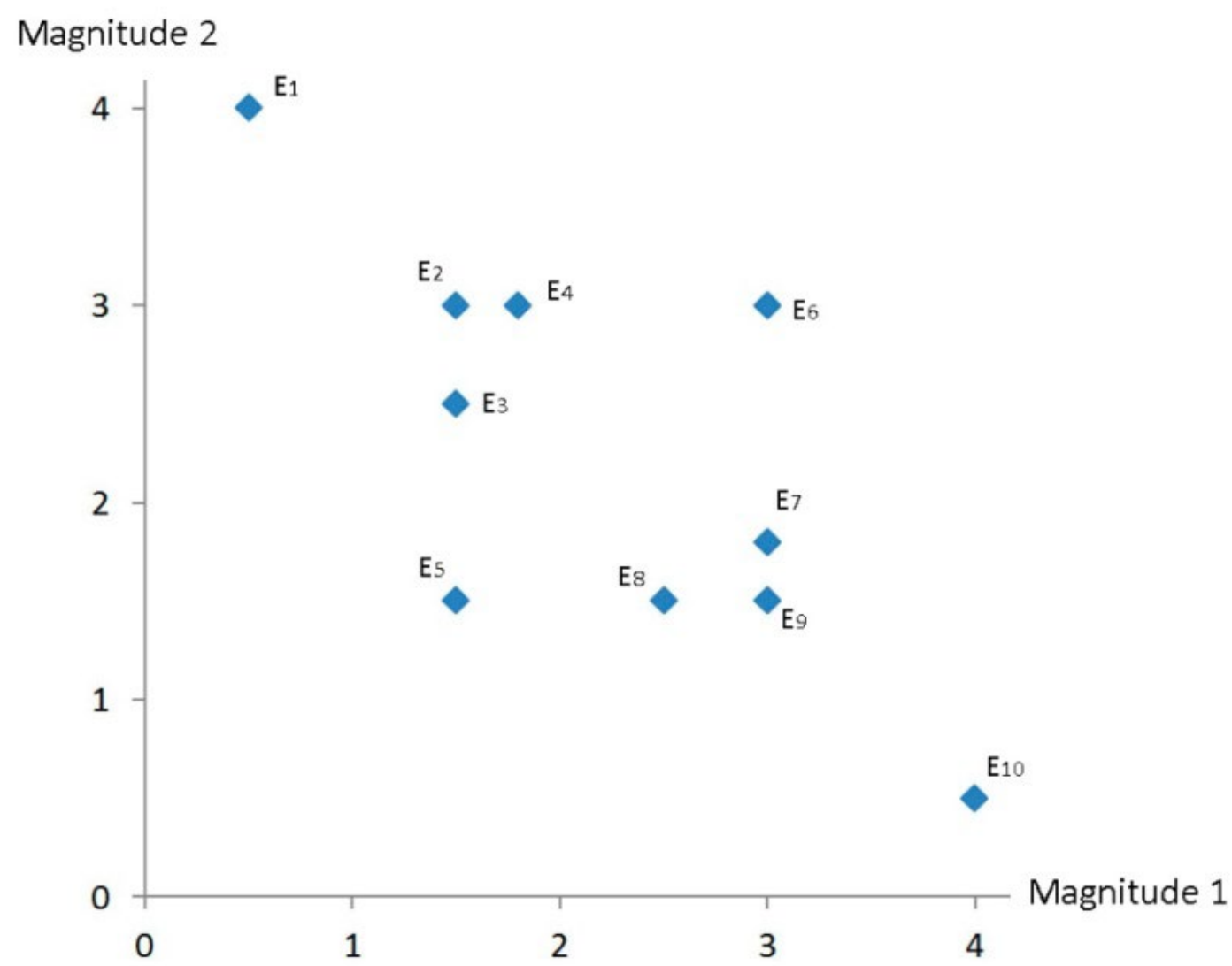

Fig. 1. Graphical representation of the ten firms in Table 1 ($E_1$ to $E_{10}$).

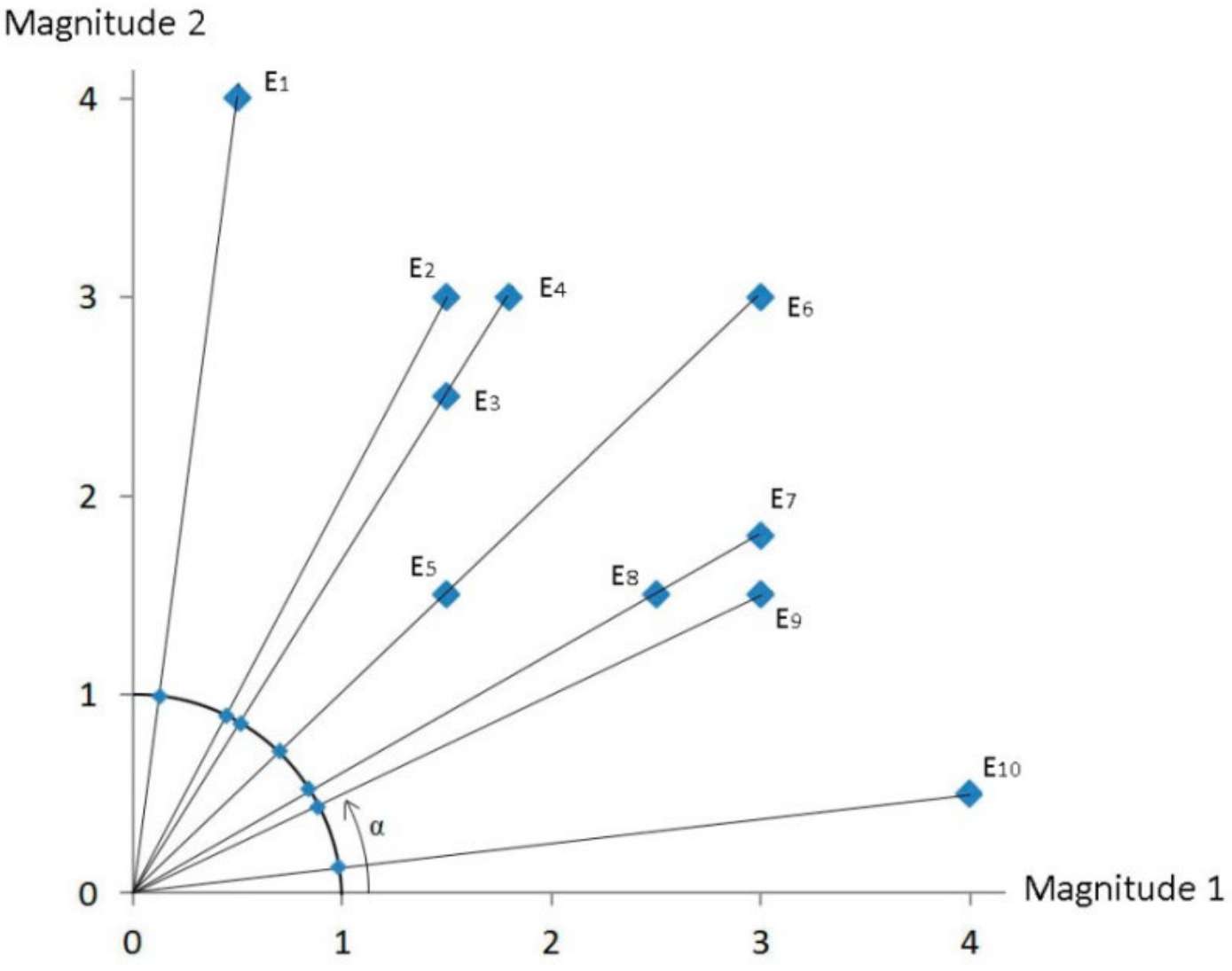

Fig. 2. Angles (α) of the ten firms ($E_1$ to $E_{10}$).

The column in Table 1 headed "Mg2/Mg1" contains the values of the ratio Magnitude 2/Magnitude 1. The same ratio values in this column for firms 3 and 4 and firms 7 and 8 are justified by firms with proportional magnitudes having the same ratio.

Geometrically, the ratio Magnitude 2 / Magnitude 1 can be interpreted as the tangent of the angle α formed between the abscissa axis and the ray that starts at the origin of coordinates and joins all the points with the same ratio value, as can be observed in Figure 3.

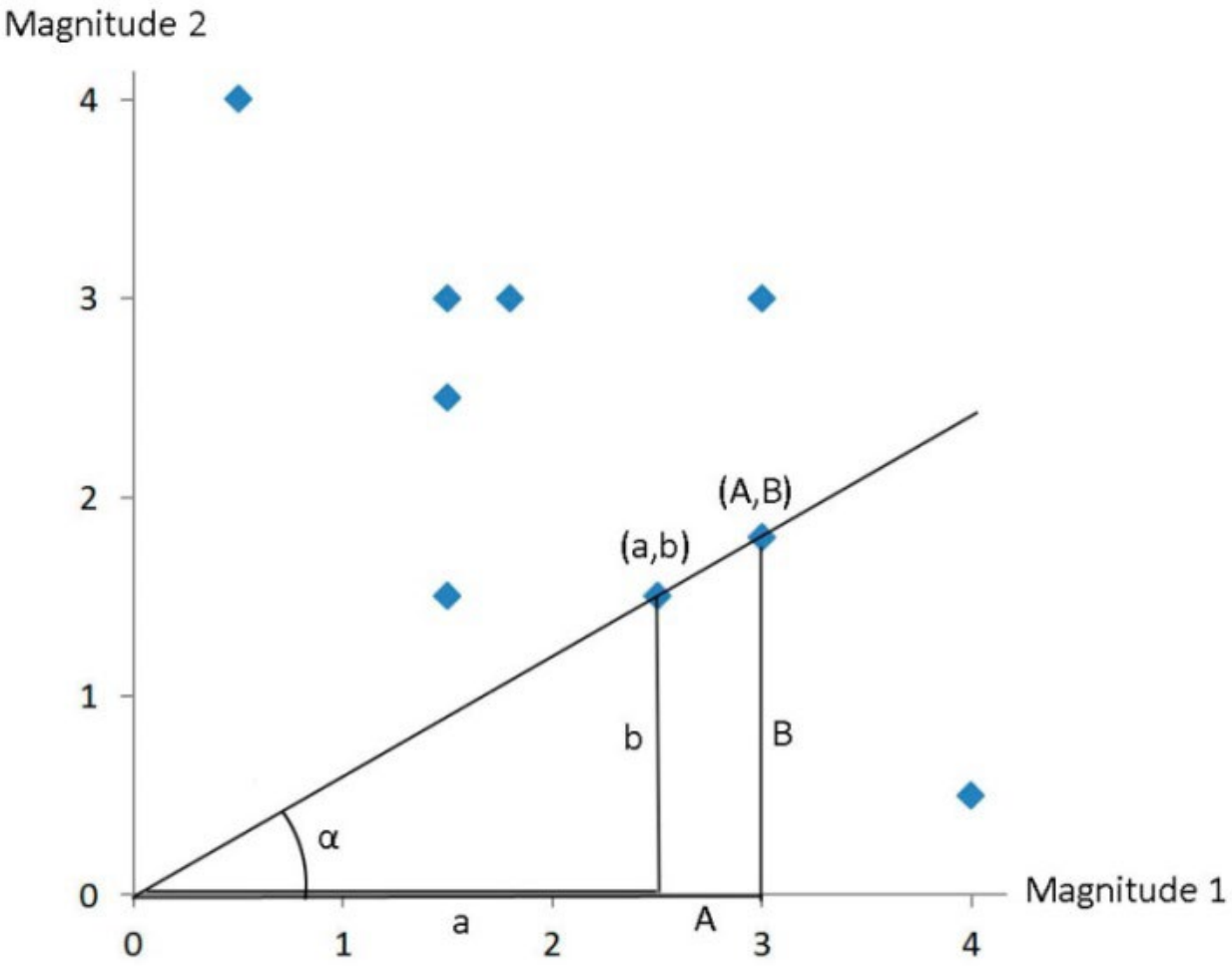

Fig. 3. Graphical relationship between the tangent of an angle and the different points on the ray that starts from the origin of coordinates (tan(α) can be written as $\frac{b}{a}$ or $\frac{B}{A}$ ).

Consequently, given that two points $(a,b)$ and $(A,B)$ that share the same ray starting from the origin of coordinates always fulfil the expression $b/a = B/A$, in the case that $a = 1$, $b = B/A$. This fact shows that the ratio Magnitude 2 / Magnitude 1 can be interpreted geometrically as the height of the cut-off point of the ray with the line of equation $x$=1, as shown in Figure 4.

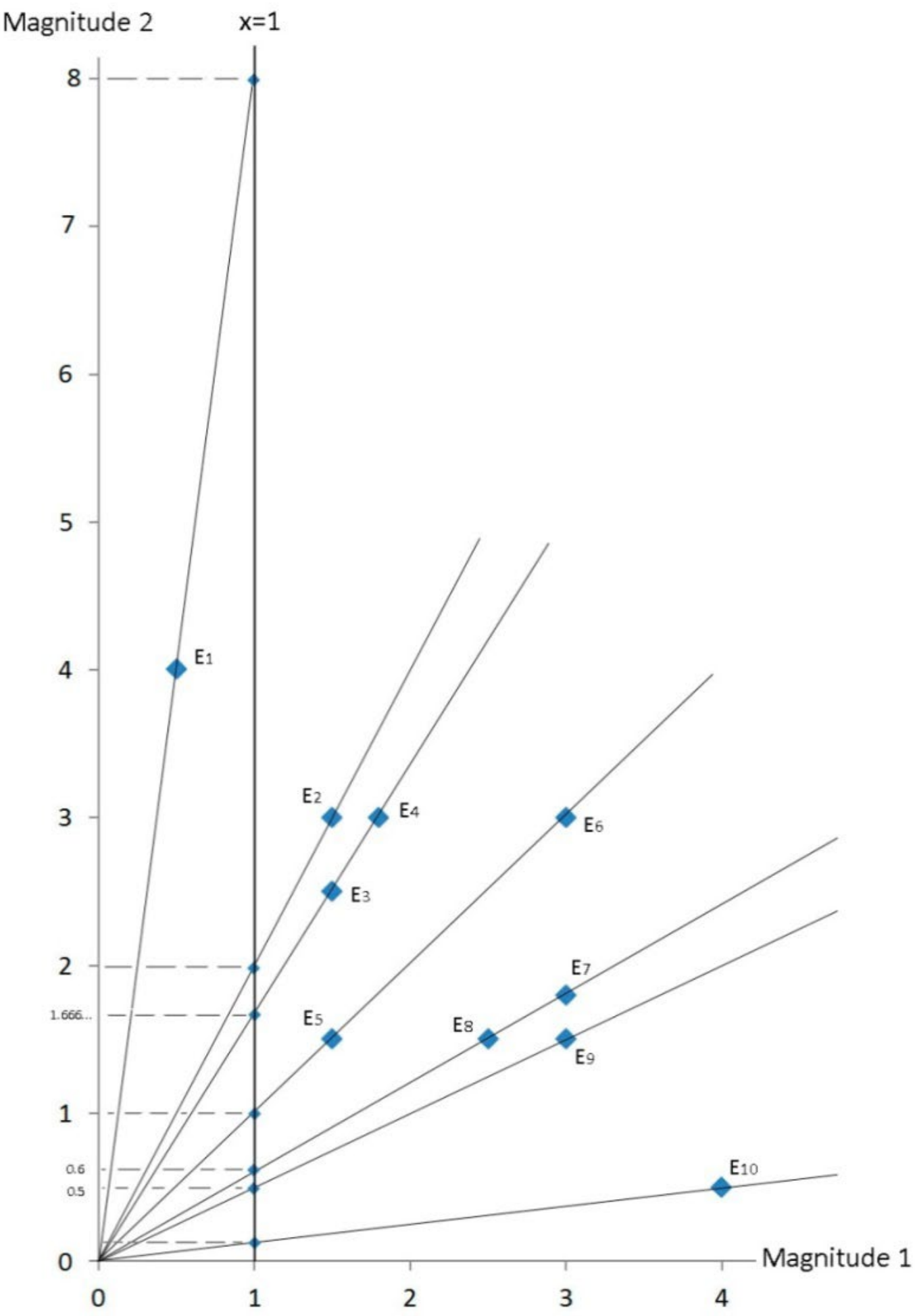

Fig. 4. Graphical calculation of the ratio of the ten firms ($E_1$ to $E_{10}$) using the projection on the line $x$=1.

Figure 4 is a clear example of the fact that the values of the ratios of a set of magnitudes with symmetry with respect to the bisector of the first quadrant are not symmetrical within the interval (0, +∞), but are distributed in a way that makes us seriously question the validity of the results obtained in sectoral studies that have employed standard financial ratios with statistical methods.

First, the tendency to obtain positive skewness in the calculation of standard ratios can be understood immediately. The fact that the points located below the bisector of the first quadrant have a ratio value that falls within the interval (0,1), and that the points located above the bisector have a ratio value that falls within the interval (1,+∞), means that above 1 the distribution of the ratio value necessarily has a longer tail.

Second, it is shown that the ratios do not preserve distances. Figure 4 shows that the distance between the ratios of the two magnitudes of firms 1 and 2 is much greater than the distance between the ratios of the magnitudes of firms 2 and 10, which is not consistent with the configuration of the data where the distance between the points on the graph representing the magnitudes between firms 1 and 2 is much shorter than the distance between the points representing the magnitudes between firms 2 and 10. This distortion of the distance produced by the ratio projection, proven with this simple example, shows that the results of any analysis using distances with ratios, such as cluster type analyses, cannot be considered as valid.

Third, the appearance of ratio values that are apparently very far from the rest, as can be seen with firm 1, is a further indication of potential distortion: firm 1 can be identified as an outlier, while firm 10 is not, in spite of the fact that its relative position with respect to the other firms is symmetrically the same as that of firm 1. In brief, there is no reason to consider firm 1 more outlying than firm 10. The appearance of such spurious outliers is a serious practical problem when using standard ratios in statistical studies. On the one hand, the elimination of firm data erroneously identified as an outlier can jeopardise the representativity of the data sample, while on the other hand, not eliminating these firms, which are in fact outliers in the value of the standard ratio, is known to distort the result of most standard statistical techniques, including regression analysis, ANOVA, and even a simple calculation of sectoral means.

Related to the previous point, there is another consideration related to the inconsistency of the results obtained when using the inverse ratio, interchanging the numerator and the denominator. From a mathematical standpoint, given two accounting magnitudes that are not zero, there is no reason beyond agreement to justify the choice as to which is the numerator and which is the denominator of the ratio, because the fact that the economic significance related to the times that the first magnitude contains the second which can be conferred on the value of a ratio is mathematically equivalent in the case of selecting the inverse ratio. To give an example, if we have the magnitudes 8 and 4, the value of the ratio 8/4 = 2 can be interpreted as the first magnitude being double that of the second, just as the value of the ratio 4/8 = 0.5 can be interpreted as the second magnitude being half of the first. In short, the same values should convey the same information. Unfortunately, this is not the case. Table 1 also contains the value of the ratio Magnitude 1 / Magnitude 2, whose values can be interpreted geometrically as the abscissa of the cut-off point of the ray with the line of equation $y$=1, as shown in Figure 5. The data of the firm identifiable as a possible outlier are, for this ratio, those of firm 10 rather than firm 1, leading us to consider that two researchers starting with the same firm data can obtain different results in their studies because they eliminate the data of different firms based on the fact that different spurious outliers appear. Even when outliers do not appear, the permutation of the numerator and the denominator modifies the distance between firms. With reference to Figure 5, the distance between the ratios of firms 1 and 2 is much smaller than the distance between the ratios of firms 2 and 10. Their skewness is also altered, and the firms previously located on the longest tail of the distribution are now located on the short one, and vice versa. This fact has special practical relevance because in the accounting literature it is quite common to encounter studies using inverse versions of the same ratio (Chen and Shimerda, 1981).

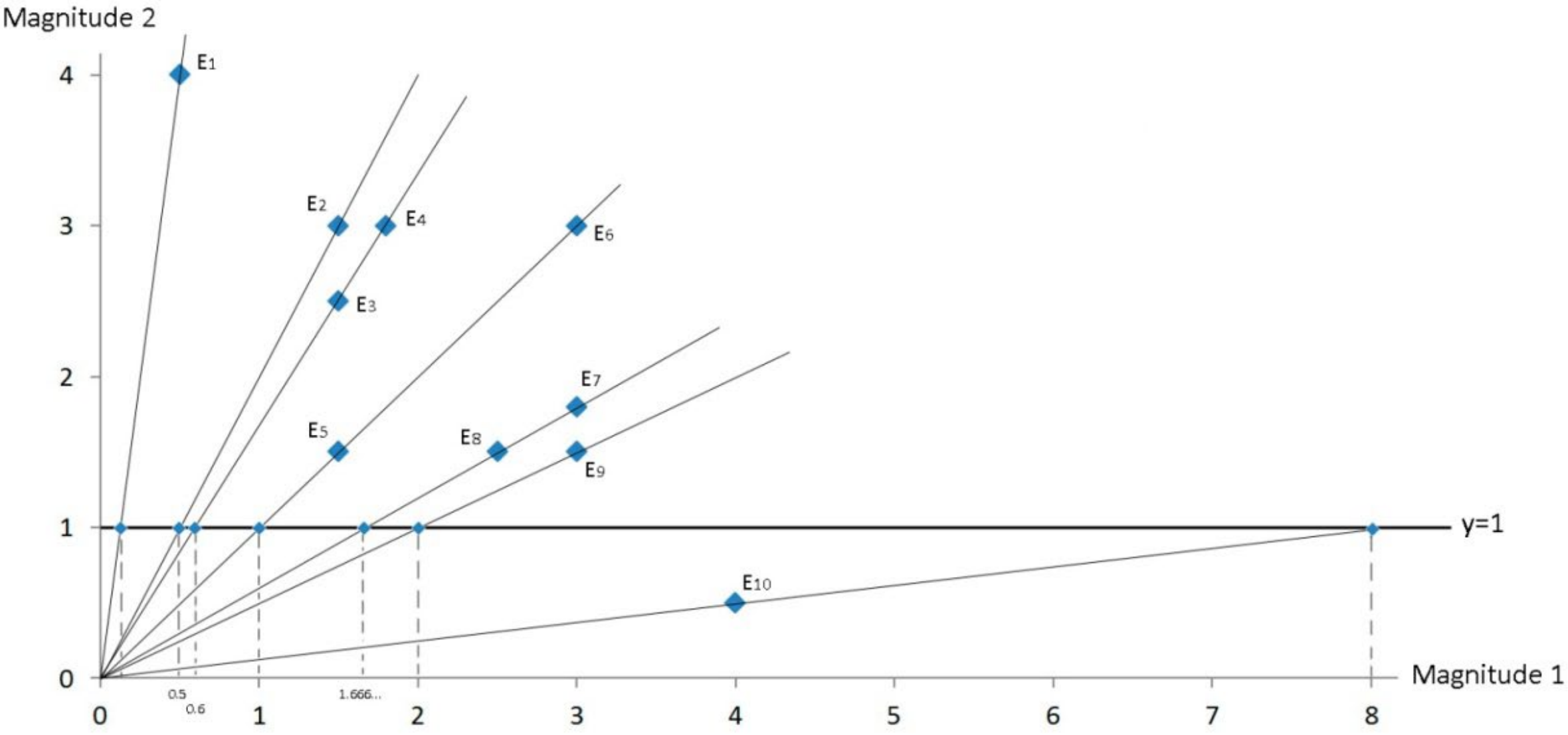

Fig. 5. Graphic calculation of the ratios of the ten firms ($E_1$ to $E_{10}$) by means of projection on the line $y$=1. Given that two points ($a$,$b$) and ($A$,$B$) that share a line from the origin of coordinates always fulfil the expression $\frac{b}{a} = \frac{B}{A}$, if $b$ = 1, then $a = \frac{B}{A}$.

## 3. Material and methods

### 3.1. Financial ratios based on the CoDa methodology

The previous section focused on showing the serious problems that can be encountered when using standard financial ratios as data in the statistical analyses of a sectoral study by using a simple mathematical counterexample. The aim of the present section is to present a methodology that solves these problems. To this end, isometric log-ratio coordinates, a type of ratio used in the so-called compositional data analysis methods, or simply speaking compositional data (CoDa), is introduced. The CoDa methodology emerged from the fields of geology and chemistry, areas in which the focus of interest of the chemical analyses carried out is the relative importance of the parts of the rock or substance being analysed, and where the size of the sample becomes irrelevant. While the CoDa methodology originated in response to the problems found when applying standard statistical methods to data of the parts of a whole often with a constant sum (Aitchison, 1986; Aitchison et al., 2000), CoDa is nowadays mainly associated with the interest in relative magnitudes, to the point where Egozcue and Pawlowsky-Glahn (2020) have defined CoDa simply as “arrays of strictly positive numbers for which ratios between them are considered to be relevant”, which fits perfectly with the analysis of financial statements using ratios.

As with standard financial ratios, the starting point when using financial ratios based on the CoDa methodology is the study objectives. For example, imagine that the aim is to review the indebtedness of firms, which would usually mean working with the following standard financial ratios:

The guarantee ratio: $r_1=x_1/(x_2+x_3)$,
the debt quality ratio: $r_2=x_2/x_3$ ,

and the following three positive-valued account categories: $x_1$ = total assets, $x_2$ = non-current liabilities, and $x_3$ = current liabilities.

Hence, the proposal of what the CoDa methodology calls isometric log-ratio coordinates (ilr coordinates) which can be interpreted as standard financial ratios to be used to study the problem in question, starts with a tree diagram, as shown in Figure 6, where each branch sequentially partitions the set of account categories, analysed in progressively smaller groups until each account category is a group in itself. At each partition of the tree, the ilr coordinates, which will herein also be referred to as compositional financial ratios, position the two groups of account categories involved in the partition (or more precisely their geometric means), one as the numerator and one at the denominator. The CoDa methodology shows that no more than two ilr coordinates will ever be needed to study the relative size of three account categories (Egozcue et al., 2003). The proposal is to use the following two compositional financial ratios to study the problem at hand:

$$y_1 = \sqrt{\frac{2}{3}}.\log\frac{x_1}{\sqrt[2]{x_2 \cdot x_3}}.$$

$$y_2 = \sqrt{\frac{1}{2}}.\log\frac{x_2}{x_3}.$$

It can be observed that the first ratio $y_1$ contains all the account categories. A scaling factor appears, multiplying the logarithm in which the total number of account categories that intervene in the ratio (2+1) appears in its denominator and, in its numerator, the product of the number account categories that appear in the denominator and the numerator (2×1). After that, the numerator and the denominator of the log-ratio are simply the geometric means of the associated components. Also note that in the numerator and the denominator the assets (numerator) and liabilities (denominator) are separated, thus providing a notion of long-term solvency, as is the case with the guarantee ratio $r_1$.

Second, the ratio $y_2$ contains the two grouped account categories in the tree diagram shown in Figure 6. The scaling factor once again contains the total number of components (1+1) in its denominator and the product of the number of account categories that appear in the denominator and the numerator (1×1) in its numerator. Note that this ratio can be interpreted as a measure of the debt quality, comparing non-current and current liability, as is the case with $r_2$.

The ratio proposal presented enables us to deduce that an ilr coordinate with a positive sign indicates that the components in the numerator are more important than the components in the denominator. From the manner in which the account categories are placed in the numerator and the denominator, we can see that a higher $y_1$ compositional financial ratio is interpreted as higher long-term solvency and a higher $y_2$ compositional financial ratio is interpreted as higher debt quality.

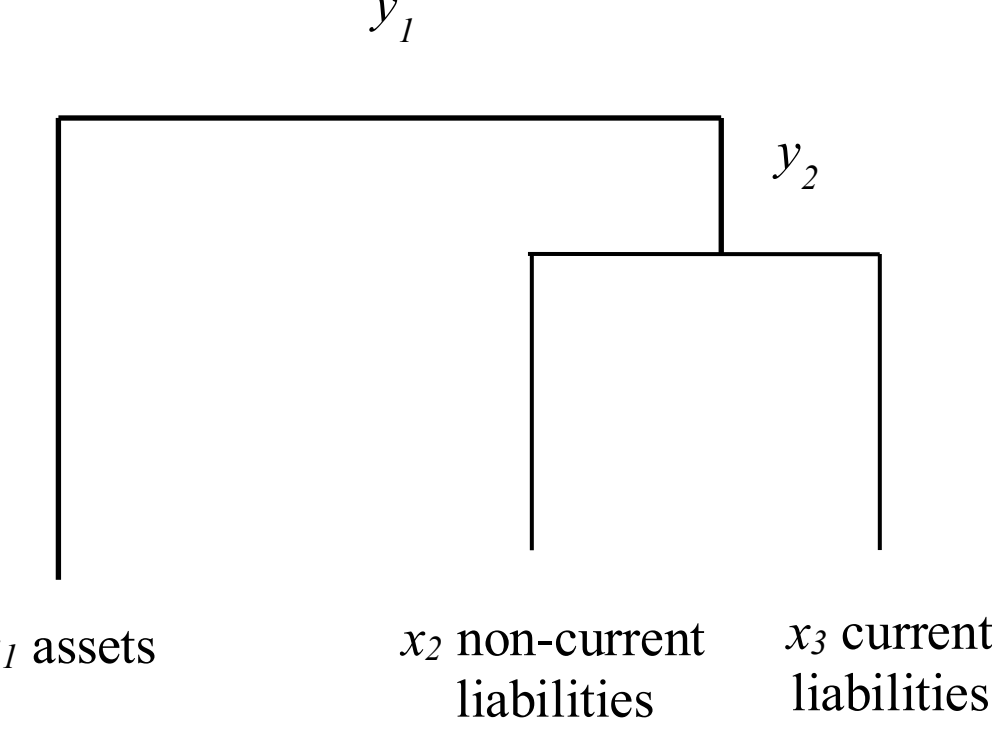

Fig. 6. Tree diagram.

This section ends with some important observations associated with the use of ratios in the CoDa methodology.

First, it must be pointed out that while any partition tree will do the job, statistically speaking, it is advisable to pick one that can be interpreted according to the usual concepts in financial ratio analysis. The tree in Figure 6 could be different, but the choice made here is aimed at obtaining two ratios related to the concepts of interest: indebtedness and debt quality, with a similar interpretation as the corresponding standard financial ratios.

Second, the choice regarding placement in the numerator or the denominator of the compositional financial ratio does not modify any other property of the log-ratio coordinate except for the sign. For instance:

$$\sqrt{\frac{2}{3}}.\log\frac{x_1}{\sqrt[2]{x_2 \cdot x_3}} = -\sqrt{\frac{2}{3}}.\log\frac{\sqrt[2]{x_2 \cdot x_3}}{x_1}.$$

This fact means that the permutation of the account categories of the numerator and the denominator ensures that:

- the same values identifiable as outliers in both cases are obtained,
- the same skewness statistic is obtained, but with the opposite sign,
- the relationships with non-financial external variables (e.g., differences of means, correlations, coefficients of regression) are identical in size, but with opposite signs.

Third, it must be pointed out that any other possible proposal of ilr coordinates can be created from linear combinations of the previous ratios. For example, an ilr coordinate that compares the assets with the non-current liability is given by:

$$\sqrt{\frac{1}{2}}\left(\sqrt{\frac{3}{2}}y_1 - \sqrt{\frac{1}{2}}y_2\right) = \sqrt{\frac{1}{2}}\log\left(\frac{x_1}{x_2}\right).$$

This observation leads us to highlight the fact that in CoDa it is not necessary to look for new ratios that improve the result presented because given a set of *D* account categories, the construction of a set of *D*-1 ilr coordinates already contains all the possible information (Egozcue et al., 2003), and adding more coordinates will not provide more. Consequently, choosing to work with compositional financial ratios has a secondary effect, eliminating the ratio mutual redundancy encountered in studies with a lot of standard financial ratios (Chen and Shimerda, 1981).

Fourth, it can be deduced from working with logarithms that the ratios have values within the entire interval $(-\infty,+\infty)$, an important fact given that the normal statistical distribution also has values within this interval.

### 3.2. Example data

The aim of the present section is to use a real example to show the different results that can be obtained from a financial statement analysis of the wine sector using standard and compositional financial ratios. The fact of obtaining different results for the same practical case study must serve to make researchers pay attention to the methodology to be followed, and to generate debate on the subject. The results of a financial statement analysis at the sector level depend on the methodology chosen and, therefore, we are of the opinion that the new methodology presented in this paper can increase the accuracy of the solvency studies of a particular sector. To this effect, this study supports selecting the new CoDa methodology proposed, the higher validity of which is shown by means of the example which follows.

Firms in the wine sector have previously developed with a clear rising tendency in terms of growth, despite some fluctuations (Martínez-Carrión and Medina-Albaladejo, 2010; Giacosa et al., 2014; Bresciani et al, 2016; Morrison and Rabellotti, 2017; Bonn et al., 2018; Merli et al., 2018; Correia et al., 2019; Arimany-Serrat and Farreras-Noguer, 2020). The wine sector is currently facing unparalleled challenges (Larreina et al., 2011; Arimany-Serrat et al., 2014; Contini et al., 2015; Sellers-Rubio et al., 2016; Sellers and Sottini, 2016; Arimany-Serrat and Farreras-Noguer, 2018) deriving from restructurings, innovations, and increasing competition among the sector's SMEs, in addition to changes in consumer habits, the price of grapes, the size of vineyards, subsidies, and the right to plant vines, which inevitably affect its financial health (Sumner, 2014; Delord et al., 2015; Lorenzo et al., 2018; Urso et al., 2018; Arimany-Serrat and Farreras-Noguer, 2020). Obtaining an accurate diagnosis of the financial health of the wine sector is therefore essential to make effective decisions regarding productions, processes, and exports to improve the sector's competitivity and ensure its survival. Furthermore, accurate knowledge of the real situation of the sector enables economic and financial decisions that directly or indirectly affect wine cooperatives, firms, and organisations to be promoted in European agendas. To this effect, the managers responsible must be aware of the need to obtain a totally accurate financial statement analysis to effectively increase the competitivity of the wine sector, a key sector in the economy of different areas of the world (Gómez-Limón Rodríguez et al., 2003; Gómez-Bezares and Larreina, 2009; Sellers-Rubio and Más-Ruiz 2013; Arimany-Serrat et al., 2016; Arimany-Serrat and Farreras-Noguer, 2020).

The data used are those of the financial situation of the Spanish wine sector, obtained from the SABI (Iberian Balance sheet Analysis System) database, developed by INFORMA D&B in collaboration with Bureau van Dijk. The search criteria were winery firms in Spain with available data for 2016 from which a sample was selected ($n$=110). We also considered the categorical variable indicating whether the firm sold at least some products using its own brand.

## 4. Results

The standard and compositional financial ratios used are those presented in the previous section, $r_1$, $r_2$, $y_1$ and $y_2$, in addition to the ratios resulting from the permutation of the numerator and the denominator, specifically $r_{1p}=(x_2+x_3)/x_1$, $r_{2p}=x_3/x_2$, $y_{1p}=\sqrt{2/3}\cdot\log(\sqrt{x_2 \cdot x_3}\ /\ x_1)$, and $y_{2p}=\sqrt{1/2}\cdot\log(x_3\ /x_2)$. These inverted ratios contain the same information as the original ones ($r_1$, $r_2$, $y_1$ and $y_2$), but they are now expressed in terms of indebtedness rather than solvency, and low quality of the debt rather than high quality.

The study focuses on the comparability of graphical displays (box plots), skewness and kurtosis, identifying any outliers present, and on the relationships between the financial statements and an external variable. In this case, the external variable is categorical, and analysis of the relationships consists in comparing the means of two groups of firms defined by the external variable using standard two-sample *t*-tests. It can also be understood as a regression analysis with a binary dummy regressor.

The box plots in Figure 7 and Figure 8 show $r_1$, $r_2$ and $r_{2p}$ to have extreme asymmetry and extreme outliers. This is not the case for $y_1$, $y_2$, $y_{1p}$ and $y_{2p}$. The graphs for $y_{1p}$ and $y_{2p}$ are simply the graphs for $y_1$ and $y_2$ turned upside down. Contrarily, the graphs for $r_{1p}$ and $r_{2p}$ significantly change their appearance with respect to $r_1$ and $r_2$.

Table 2 once again shows $r_1$, $r_2$ and $r_{2p}$ to have extreme asymmetry and extreme outliers. The statistics and outliers for $y_{1p}$ and $y_{2p}$ are the same as for $y_2$ and $y_{1p}$ after reversing the skewness sign, and reveal far lower skewness, far fewer outliers, and very rare extreme outliers.

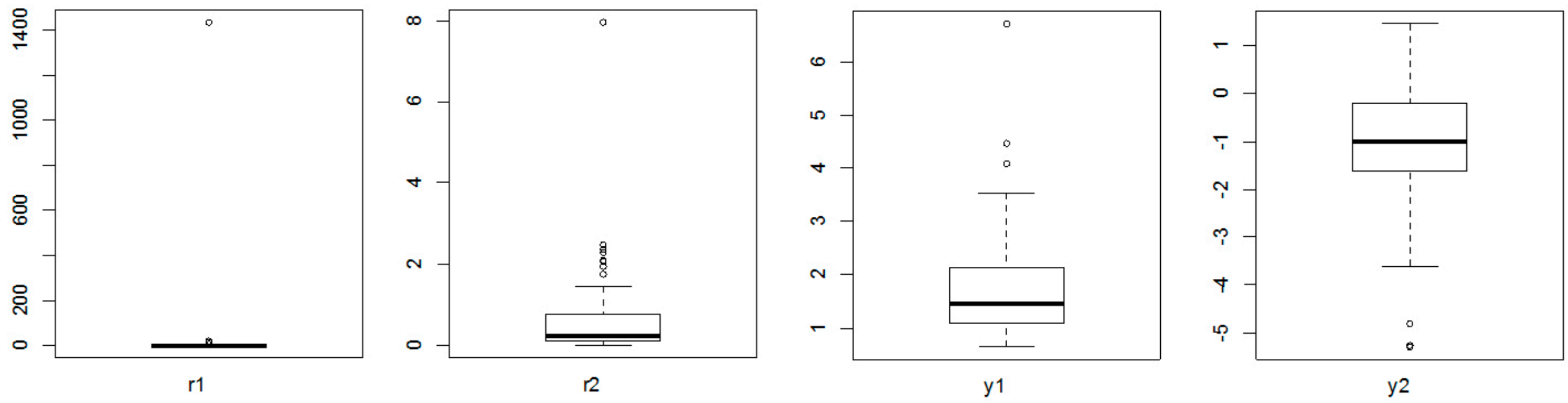

Fig. 7. Box plots of log-ratio coordinates and standard financial ratios.

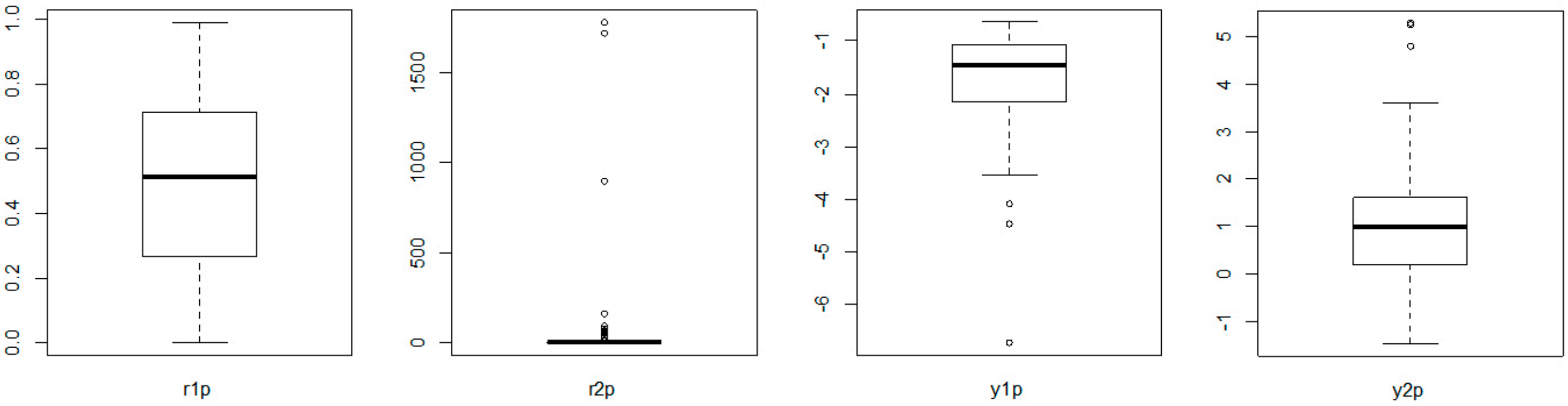

Fig. 8. Box plots of log-ratio coordinates and standard financial ratios after permutation.

Table 2.
Skewness, kurtosis, and count of outliers.

| | Skewness | Kurtosis | n outliers* | n extreme outliers** |
|---|---|---|---|---|
| $r_1$ | 10.48 | 109.87 | 10 | 7 |
| $r_2$ | 5.27 | 38.85 | 9 | 1 |
| $y_1$ | 1.97 | 6.53 | 3 | 1 |
| $y_2$ | -0.98 | 1.52 | 3 | 0 |
| $r_{1p}$ | -0.01 | -1.16 | 0 | 0 |
| $r_{2p}$ | 6.34 | 40.72 | 22 | 17 |
| $y_{1p}$ | -1.97 | 6.53 | 3 | 1 |
| $y_{2p}$ | 0.98 | 1.52 | 3 | 0 |

* beyond 1.5(Q3-Q1) from the box boundary in the box plot.
** beyond 3(Q3-Q1) from the box boundary in the box plot.

In Table 3, $y_2$ and $y_{2p}$ equivalently show firms with at least one own brand to have a significantly better debt quality. This relationship is also marginally significant for $r_2$ but not for $r_{2p}$. $r_1$ and $r_{1p}$ are also in contradiction with respect to a relationship between long-term solvency and having an own brand. This relationship does not emerge with $y_1$ and $y_{1p}$. The results regarding relationships with external variables thus change from standard to compositional financial ratios, and in the former case they also change depending on the numerator and denominator choice.

Table 3.
Equal variance Two-Sample t-test (positive t-value means “yes” group -having at least one brand- has a larger ratio mean than the “no” group), R-squared (percentage of explained variance of the ratio in a dummy variable regression with having or not an own brand as predictor).

| Ratio | t-value | p-value | R-squared |
|---|---|---|---|
| $r_1$ | 0.53 | 0.59 | 0.3% |
| $r_{1p}$ | -2.23 | 0.03** | 4.4% |
| $r_2$ | 1.88 | 0.06* | 3.2% |
| $r_{2p}$ | -0.78 | 0.44 | 0.6% |
| $y_1$ | 0.25 | 0.80 | 0.1% |
| $y_{1p}$ | -0.25 | 0.80 | 0.1% |
| $y_2$ | 2.14 | 0.03** | 4.1% |
| $y_{2p}$ | -2.14 | 0.03** | 4.1% |

* Significant at 10%
** Significant at 5%

As expected, we find that log-ratio coordinates are better suited for statistical analysis, having fewer outliers and less skewness and kurtosis, and that permuting denominator and numerator does not modify the results. On the contrary, when using standard financial ratios, permuting the numerator and denominator leads to substantial differences in the conclusions: outliers emerge or disappear, and the relationships with the external variable become significant or insignificant.

## 5. Discussion, limitations, and future research

The main objective of this study is to show that the use of standard financial ratios in sectoral studies is not recommended because the results produced can be invalid, even when applying the most elementary statistical techniques. The present article complements the already existing research that exposes the serious consequences of using financial ratios in multivariate statistical analyses (Cowen and Hoffer, 1982; Linares-Mustarós et al., 2018; Creixans-Tenas et al., 2019; Carreras-Simó and Coenders, 2021), also revealing the possible causes of the problems. In line with what has been shown in this study, it must be remembered that because asymmetrical distributions can lead to the relations between the standard financial ratios being non-linear (Cowen and Hoffer, 1982), using the latter can impede, for example, the application of factor analysis and regression analysis. Another obvious example of the non-validity of results can be found in the specific case of cluster analysis. In asymmetric distributions, some of the clusters can end up being very small (Feranecová and Krigovská, 2016; Santis et al., 2016; Sharma et al., 2016; Yoshino et al. al., 2016), which is detrimental to the grouping of firms. This problem is related to the appearance of spurious outliers, which we have already seen, and in this case tends to generate clusters with a single firm (Linares-Mustarós et al., 2018). This problem is often ignored in practice even though it is very well-known in theory (Lev and Sunder, 1979; Cowen and Hoffer, 1982; So, 1987; Ezzamel and Mar-Molinero, 1990; Watson, 1990; Balcaen and Ooghe, 2006), causing extremely asymmetric distributions, the possible main source of which are the aforementioned spurious outliers (Frecka and Hopwood, 1983; Carreras-Simó and Coenders, 2021), a fact that could be a source of the invalidity of even any conclusions drawn from studies related to ratio averages (Linares-Mustarós et al, 2013, Rondós Casas et al., 2018).

The mathematical counterexample and the empirical example in the present study have both shown that working with financial ratios based on the CoDa methodology minimises the problem of the appearance of outliers and solves the problem of asymmetry of information in choosing the numerator and the denominator, a well-identified problem in financial statement analysis (Frecka and Hopwood, 1983).

The study has shown that the construction of two CoDa ilr coordinates related to three accounting magnitudes enables the indebtedness of the wine sector in Spain to be studied. The methodology presented can be extended to the study of any $D$ accounting magnitudes with $D$-1 ilr coordinates, such that their construction can meet any other study objectives. This opens up the possibility of enormously simplifying applied research, related to the fact that it can be shown that only $D$-1 ilr coordinates include information about all the possible ratios between any two account categories of the $D$ account categories considered (Pawlowsky-Glahn et al., 2015), thus preventing the redundancy problems encountered in the financial literature when a very large number of ratios are used (Chen and Shimerda, 1981; Barnes, 1987; Carreras-Simó and Coenders, 2020). The different ways the tree can group the $D$ magnitudes will shape the objectives of the study. We would like to highlight that, if it were required, other positive magnitudes whose size is to be compared with accounting magnitudes in relative terms could be added. This even applies to non-monetary magnitudes such as number of employees, and other magnitudes that are usual in management ratios, strategy evaluation, and performance. To this effect, the fact that there are no restrictions on the type of magnitude used must be underlined, as long as they are positive.

Another potential way to broaden this research is related to the various beneficial secondary effects produced by using financial ratios based on the CoDa methodology, for example, the improvement in the distribution of the magnitudes of the compositional ratios such that it is normal or nearly so (Aitchison, 1986; Pawlowsky-Glahn et al., 2015), a characteristic rarely seen in standard ratios (Deakin, 1976; So, 1987; Ezzamel and Mar-Molinero, 1990; Martikainen et al., 1995; Mcleay and Omar, 2000). Another effect derives from calculating the Euclidean distance over ilr coordinates, which is the same as working with the Aitchison distance over accounting magnitudes (Aitchison, 1983; Egozcue et al., 2003; Linares-Mustarós et al., 2018). This implies not having to modify computer programs to include new distances when, for example, making classifications, the calculations for which can be made with the Euclidean distance over the transformations of these accounting magnitudes in the form of ilr coordinates. Re-visiting current financial statement analysis studies using the CoDa methodology could be a source of multiple related papers.

One limitation to bear in mind when working with the CoDa methodology, which is the same for standard financial ratios, derives from the mathematical assumption that ratios whose components can simultaneously be zero and positive and negative make no sense. First, a ratio with a zero numerator has no possible interpretation, so it cannot be deduced from the zero value that the numerator is 10 times smaller than the denominator, or 100 or 1000. The zero value of this ratio simply contributes no information. If the denominator is zero we cannot even compute the ratio. The case of simultaneous positive and negative magnitudes is even more serious. It is unjustifiable that two firms with close values offer interpretations that make us think that these very similar firms are completely different. More specifically, if we have two almost identical firms with the same numerator, but with denominators that are practically 0, one slightly positive and the other slightly negative, their values for the ratio can be as different as one wishes. Hence, comparing the values of this ratio is meaningless. What comparative information about financial health would be extracted from a value of thousand for the first ratio, and minus two million for the second ratio? Is the first firm so much better off in terms of financial health than the second? In this line, Stevens (1946) already pointed out that computing a ratio is a meaningful operation only for variables in a ratio scale, which need to have a meaningful absolute zero and thus no negative values. Unfortunately, magnitudes like profit, cash flow, net worth and working capital, which can be negative and dispense with this recommendation, have become usual in financial ratios. In the CoDa methodology, the assumption of non-negative magnitudes is an unnegotiable assumption, which is why avoiding negative components is recommended, converting them into positive components, which can always be done without losing any information. For example, replacing the ratio of profits over assets with two strictly positive ratios, one of income over assets and the other of expenses over assets, is recommended. The difference between the two is the ratio replaced. In the same vein, ratios computed from current assets and current liabilities should replace ratios computed from the working capital. This recommendation has already been made in the area of accounting (Lev 1969; Lev and Sunder, 1979) but has unfortunately generally been overlooked. Regarding the zero-value issue, the CoDa methodology provides a diverse toolbox with imputation methods under the most common assumptions (Palarea-Albaladejo and Martín-Fernández, 2015).

Last, it must be mentioned that in line with other CoDa methodology proposals, apart from the ilr coordinates, other types of compositional financial ratios can be used to replace standard financial ratios. These include pairwise log-ratios (Creixans-Tenas et al., 2019) and centred log-ratios (Carreras-Simó and Coenders, 2020; Saus-Sala et al., 2021). The advantage of the ilr coordinates used in this paper is their general applicability to any statistical method (Linares-Mustarós et al., 2018; Carreras-Simó and Coenders, 2021; Arimany-Serrat et al., 2022). For example, pairwise log-ratios are not applicable to statistical methods that use distances, such as cluster analysis, and centred log-ratios are not applicable to statistical methods that require the inversion of covariance matrices, such as multivariate analysis of variance (MANOVA).

## 6. Conclusions

This article was developed around the following three key ideas.

First, the study involved a detailed analysis of the fact that while it is a valid methodology for studying the financial reality of single organisations, the methodology of standard financial ratios presents serious problems when the ratios are used as variables in sectoral statistical analyses, and can produce, for example, average values contaminated by spurious outliers or data bounded to be positive which can never follow a normal statistical distribution.

Second, the study showed that the cause of the abovementioned problems and other related problems is the asymmetry produced in calculating standard financial ratios. The fact of dividing two numerical values causes a distortion of the symmetry per se. A simple mathematical counterexample was enough to evidence this fact.

Third, a new working methodology was presented, which enables the problems associated with the use of standard financial ratios to be eliminated or minimised. The methodology is based on a type of ratio used in the so-called CoDa methods. The soundness of these methods has been shown in many theoretical works, offering the hope that its applicability to accounting science can be a possible source to solve current methodological problems.

These three key ideas open up a dialogue in the area of accounting around the need to find a new working methodology to diagnose the financial health of a business activity. Given that more reliable results provide a better basis for taking economic and financial decisions, the search for a reliable methodology must be a current interest for the accounting scientific community. To this effect, the present work takes the wine sector as an example for comparative studies among different methodologies to diagnose financial health in the different activity sectors to decide what the most suitable policies to create added value are. We hope, therefore, that this will encourage similar studies focused on the different business sectors and aimed at obtaining a more accurate financial statement analysis.

## Conflicts of interest

There are no conflicts of interest.

## Acknowledgements

This research was supported by the Spanish Ministry of Science and Innovation/AEI/10.13039/501100011033 and by ERDF A way of making Europe (grant PID2021-123833OB-I00), the Spanish Ministry of Health (grant CIBERCB06/02/1002), and the Government of Catalonia (grants 2017SGR656, 2017SGR386 and 2017SGR155).